# Multipath Amplification of Chaotic Radio Pulses and UWB Communications

Yuri V. Andreyev, Alexander S. Dmitriev, *Member, IEEE*, Andrey V. Kletsov

*Abstract* — Effect of multipath amplification is found in ultrawideband wireless communication systems with chaotic carrier, whereas information is transmitted with chaotic radio pulses. This effect is observed in multipath environment (residence, office, industrial or other indoor space). It is exhibited as an increase of signal power at the receiver input with respect to the case of free space. Multipath amplification effect gives 5–15 dB energy gain (depending on the environment), which allows to have 2–6 times longer distance range for the same transmitter power.

## I. INTRODUCTION

Recently, ultrawideband (UWB) signals were introduced in practice of public communications. As a rule, UWB signals are defined as signals with relative bandwidth $2\Delta f/(f_u + f_l) > 0.25$, where $\Delta f = f_u - f_l$, and $f_u$ and $f_l$ are upper and lower boundaries of the signal frequency band [1, 2]. Sometimes, different definitions are used, e.g., signals with bandwidth $\Delta f > 500$ MHz [3].

UWB signals can be implemented with ultrashort pulses [1, 2, 4, 5], chaotic oscillations [6–9], OFDM modulation [10], chirps [11].

Since 2002, some countries permitted unlicensed use of UWB communications in 3–10 GHz range, many others are in process [3, 12, 13]. In order to repeatedly use the already occupied frequency range and to avoid interference with existing and future narrowband communications systems (e.g., Wi-Fi, Wi-Max, etc.), permitted spectral density of UWB wireless communications in 3–10 GHz range is limited, e.g., in USA it is $S(f) \leq -41.3$ dBm/MHz. Thus, total emitted power in the entire permitted band must not exceed −2.3 dBm (approx. 600 μW). With omni-directional antennae these restrictions confine the UWB signal application area to wireless communications at distances 1…10 (100) m.

Majority of tasks to be solved by corresponding UWB communications systems occur in residence, office and industrial, i.e., in multipath indoor environment. Hence, the performance of UWB systems in multipath environment is of prime importance.

Multipath propagation can severely impair the performance of a wireless digital communication system. Typical negative effects in narrowband communication systems are fading and inter-symbol interference [14]. To compensate for these effects, sophisticated constructions (e.g., rake-receiver) were invented [14].

The behavior of a communication system in multipath environment is determined by its structure, the type of carrier signal and modulation scheme. A few years ago a concept of direct chaotic communications (DCC) was proposed [6–9]. This concept is described in Section 2. IEEE Standard Committee included direct chaotic systems in IEEE 802.15.4a standard (as option). In this paper we theoretically consider operation of DCC systems in multipath channel and show that for such systems there is a unique effect which we call "multipath amplification", and which at certain system parameters exhibits itself as an increase of average signal power at the output of the receiver with respect to «free-space» case.

In Section 3 the nature of multipath amplification effect is discussed. To simulate operation of a communication system in multipath environment, multipath channel models are necessary. These models are presented in Section 4, and estimates of multipath gain are obtained in Section 5. In Section 6 conditions for this effect are summarized. Effect of multipath amplification on the range of communication system with chaotic radio pulses (for IEEE 802.15.4a standard) is evaluated in Section 7.

## II. DCC TECHNOLOGY

The technology of direct chaotic communications is based on the use of chaotic signals, which are irregular oscillations produced by deterministic systems [15]. Irregular behavior of such systems is not a consequence of random factors or components, but is caused by inner dynamics. As a rule, practical structure of chaotic generators is simple, yet with qualified design they produce chaotic oscillations in the required frequency band with good efficiency (up to 20–30%) and even with the required spectrum shape [16, 17]. Chaotic generators should be treated as special effective source of noise-like oscillations.

The technology of direct chaotic communications employs such attractive features of chaotic signals as naturally wide bandwidth and relative simplicity of communication systems with chaotic signals.

A key notion of the proposed technology is chaotic radio pulse, which is a fragment of signal with duration essentially larger than quasi-period of chaotic oscillations. The bandwidth of chaotic radio pulse is determined by the bandwidth of original chaotic signal, and it is practically independent of the pulse duration (in a wide range of pulse duration variation). This makes chaotic radio pulse essentially different from the "classical" radio pulse filled with narrowband periodic signal, whose bandwidth is determined by pulse duration.

The technology of direct chaotic communications is based on the following three ideas:

(а) chaos source produces chaotic oscillations directly in RF communication channel band;

б) information is put in the chaotic signal by means of forming corresponding sequence of chaotic radio pulses;

в) information is retrieved from RF signal without intermediate frequency conversions.

The structure of a DCC system is shown in Fig. 1.

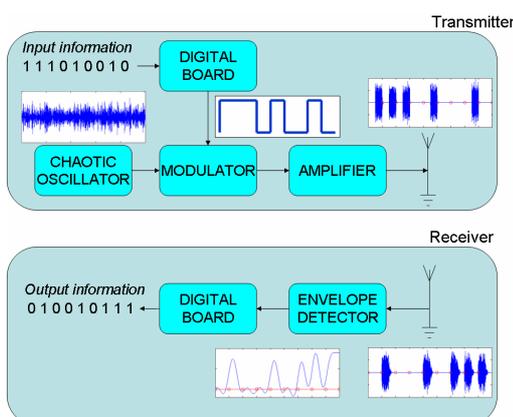

Fig. 1. Structure of direct chaotic communication system: transmitter with external modulation and receiver

Naturally wide bandwidth of chaotic signals gives simplicity of signal generation and modulation as compared with ordinary spread spectrum systems. Since the spectrum of DCC carrier signal is independent of the pulse duration and, consequently, of the signal base $B = \Delta f \cdot T_S$, where $\Delta f$ is signal bandwidth and $T_S$ is pulse duration, DCC has no energy restrictions characteristic of the systems with short pulses. Here, energy per bit $E_b$ can be varied in wide range by means of varying the pulse duration. So, with DCC one can implement the "one symbol – one pulse" idea in a wide range of transmission rates. Incoherent receiver of chaotic signals (envelope detector) can be easily realized with less component count (without mixers and PLLs) than alternative solutions.

An example of implemented DCC transceiver is shown in Fig. 2. Transceiver PPS-40 is intended as communication device of UWB wireless sensor networks. It has several external interfaces (UART, SPI, I2C) and two analog channels. Various devices can be connected, such as sensors, sources of audio or video signals, etc. Transceiver's firmware can be easily modified to satisfy a specific application.

Physical layer of this device corresponds to the optional solution with chaotic signals of IEEE 802.15.4a standard [12]. The transceiver printed-circuit board (PCB) carries a UWB antenna, RF switch, a generator of chaotic pulses, an envelope detector with 60-dB dynamic range (a quadratic detector with low-pass filter), a system for digital processing of analog signal implemented on FPGA, and a microcontroller responsible for MAC layer. Information is transmitted and received in packets. The packet structure is determined by the standard.

*Device parameters:*
- frequency band 3.1…5.1 GHz;
- average (in time) spectral density below –41.3 dBm/MHz;
- average emitted power (rate 2.5 Mbps) –10 dBm;
- average emitted power (rate 0.1 Mbps) –24 dBm;
- distance 10…12 m;
- maximum PHY rate of transmission/reception 2.5/2.5 Mbps;
- power supply voltage 3.5…4.5 V.

Data on power consumption are given below.

*Transmission mode:*
- RF block supply current 0.4 mA (100 Kbps), 4 mA (1000 Kbps)
- Digital block supply current 5 mA (100 Kbps), 30 mA (1000 Kbps)

*Reception mode:*
- RF block supply current 40 mA (100 and 1000 Kbps)
- Digital block supply current 5 mA (100 Kbps), 30 mA (1000 Kbps)

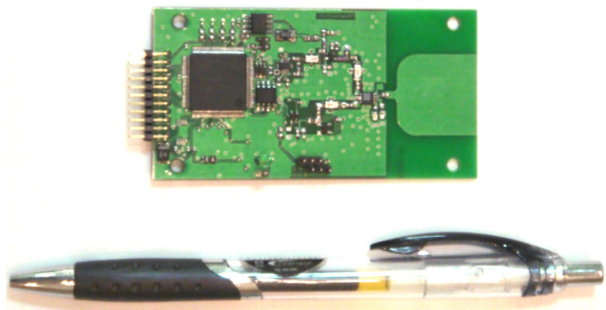

Fig. 2. PCB of UWB DCC transceiver PPS-40

DCC systems can use various types of modulation: presence/absence of chaotic pulse on prescribed position (chaotic on-off keying, COOK), differential chaotic shift keying (DCSK), pulse position modulation (PPM), etc. Since information is transmitted by a pulse stream, not only modulation type is important, but also the pulse duration and duty cycle. These parameters define the rate of communication system and its immunity in various types of communication channels.

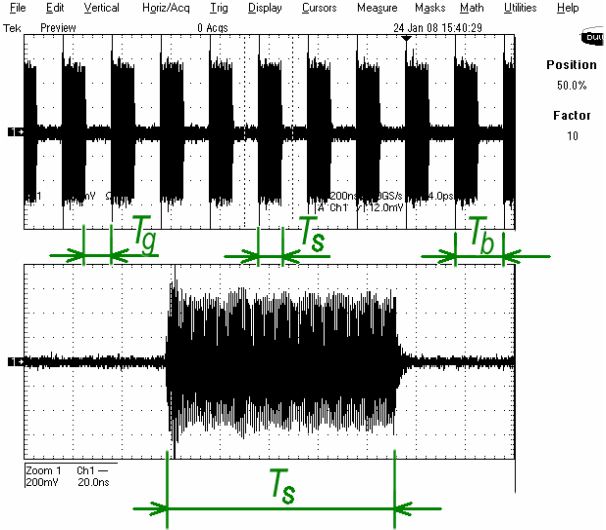

Fig. 3. Chaotic radio pulse train

In direct chaotic communication (DCC) system that is considered in this paper, one of the simplest modulation methods is used: symbol "1" is transmitted by means of emitting chaotic radio pulse of duration $T_s$ on a certain position in time domain, and "0" by means of emitting nothing on that position. Pulse positions are separated by guard intervals of duration $T_g$. Thus, information bit is transmitted on time interval $T_b = T_s + T_g$, which determines the maximum transmission rate $C = 1/T_b$ (Fig. 3).

In order to decide what symbol was transmitted on current time position, incoherent receiver collects energy on this position and compares its value with a threshold. The threshold is set at the stage of establishing connection, or it is set constant by device fabrication. Integrator that collects energy on a certain time interval is made with low-pass filter (LPF). Interval $T_b$ on which the signal power is integrated is related with the filter cutoff frequency $f_{cutoff}$ as $T_b = 1/f_{cutoff}$. Parameter $f_{cutoff}$ must be set carefully, so that the energy be collected only on interval $T_b$, not on guard interval. Otherwise, besides the signal energy the receiver would pick up noise at the receiver input, which would decrease $E_b/N_0$ (energy-per-bit to spectral-density of noise ratio) and deteriorate the receiver effectiveness. In the discussed circuit the received signal envelope goes to digital block where the decision on the received symbol is made.

Despite the simplicity of incoherent receiver, in multipath channel is appears to be close to optimal receiver, as will be shown below.

### III. AMPLIFICATION OF CHAOTIC RADIO PULSES IN MULTIPATH CHANNEL

As a rule, multipath propagation causes degradation of communication system performance. However, special receiver can substantially improve the system performance with respect to single-path propagation (Fig. 4).

Let us consider the signal after multipath channel at the receiver input. Chaotic signals that come to meeting point on different paths appear practically uncorrelated. This is due to the fact that autocorrelation time τ of chaotic signals with uniform spectrum (as well as of random signals) is inversely proportional to signal bandwidth $\Delta f$: as follows from Wiener-Khinchin theorem, $\tau \cong 1/\Delta f$ [18]. That is, if the bandwidth is $\Delta f$ = 1 GHz, then $\tau$ = 1 ns. This means that two signals coming to a meeting point on two different paths with time delays differing by more than $\tau$ are uncorrelated. If uncorrelated signals are summarized, the resulting signal energy is the sum of the energies of the added terms (signals).

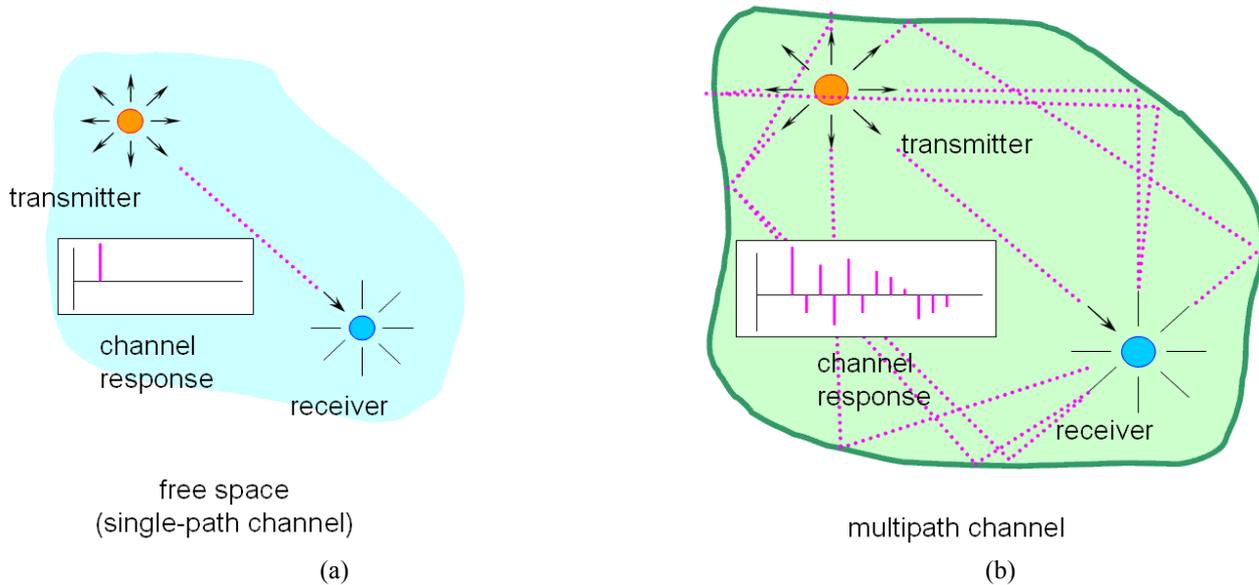

Fig. 4. (a) In single-path case (a) small part of energy is delivered to the receiver, whereas (b) in multipath case more energy comes to the receiver input

Good receiver must be capable of collecting the energy of signals coming from different paths. This can be effectively solved by one of the simplest circuits, i.e., by envelope detector composed of a quadratic detector and LPF. Indeed, mathematically the quadratic detector raises the signal to the second power, thus giving instantaneous power of the input signal (which is the sum of signals from different paths). LPF averages the power signal and effectively eliminates cross-correlation components.

However, in the discussed communication system chaotic radio pulses (fragments of chaotic signal) are used, not the continuous chaotic signal. Adding pulses differs from adding continuous signals because it is limited in time by the pulse duration. The channel delay spread must be also taken into account. In. Fig. 5 two cases of pulse propagation through multipath channel are shown schematically: a long and a short pulse. In both cases the pulses become much longer, but if the delay spread is short compared to the pulse duration, then the most part of the energy of delayed paths hits the pulse position $T_b$, so that one can say that the paths' energies are summed (Fig. 5a). Otherwise, if pulses are short compared with the delay spread and especially compared with autocorrelation time $\tau$, then pulses coming on different path do not overlap at the receiver input and they cannot add energy to the main pulse (Fig. 5b).

In DCC system, the signal in the receiver is processed as follows. Incoming signal undergoes quadratic transformation, then it is filtered in the frequency band of information signal. Thus, the received signal envelope is retrieved, which then goes to digital part of the receiver. Note, that the LPF cutoff matches the duration of original chaotic radio pulses, which means that the energy is collected only on time interval $T_S$. The rest of the energy that is spread beyond the original pulse position $T_S$ is discarded. If the receiver collected energy of the delay spread, more noise would also have been collected and the resulting $E_b/N_0$ would have become worse.

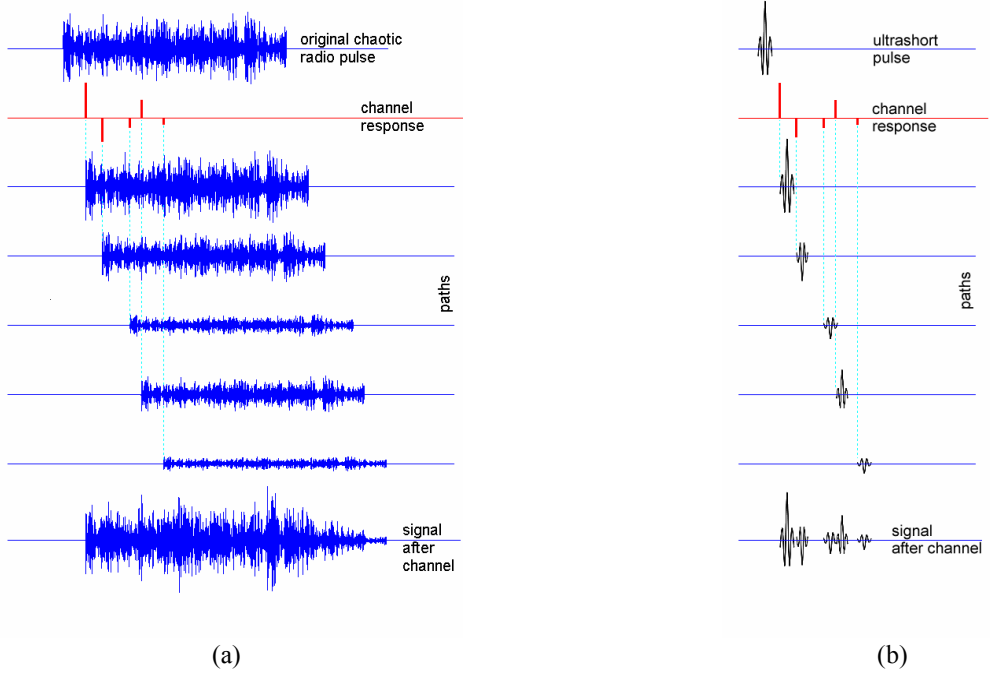

(a)                                      (b)

Fig. 5. Propagation of (a) long and (b) short pulse through multipath channel

The receiver of chaotic radio pulses can be described with the following relations (discrete-time model). Let at the receiver input there be a set of path signals $x_i(k \cdot \Delta t)$, where $\Delta t$ is time sampling interval defined by the signal bandwidth as follows from Nyquist theorem; $k$ is discrete time; $i = 1, \ldots, L$ is path index. Thus, the signal at the input of quadratic detector is equal to

$$x(k \cdot \Delta t) = \sum_{i=1}^{L} x_i(k \cdot \Delta t). \qquad (1)$$

The signal at the output of the detector at moment $(k \cdot \Delta t)$ is the instantaneous power of input signal. Knowing that the power of $i$th path signal is equal to

$$W_i(k \cdot \Delta t) = x_i^2(k \cdot \Delta t), \qquad (2)$$

at the output of quadratic detector we have signal

$$W(k \cdot \Delta t) = \left( \sum_{i=1}^{L} x_i(k \cdot \Delta t) \right)^2 =$$
$$= \sum_{i=1}^{L} x_i^2(k \cdot \Delta t) + \sum_{i \neq j}^{L} \sum_{j \neq i}^{L} x_i(k \cdot \Delta t) x_j(k \cdot \Delta t) = \qquad (3)$$
$$= \sum_{i=1}^{L} W_i(k \cdot \Delta t) + \sum_{i \neq j}^{L} \sum_{j \neq i}^{L} x_i(k \cdot \Delta t) x_j(k \cdot \Delta t).$$

Low-pass filter (LPF) after quadratic detector operates as an integrator, i.e., it averages the signal. At the output of LPF we obtain sum energy of the paths arriving at the receiver input, the energy is collected over time interval equal to characteristic time of the filter:

$$E = \sum_{n=1}^{M} W\big((k+n-1)\cdot \Delta t\big) \approx$$
$$\approx \sum_{n=1}^{M}\sum_{i=1}^{L} W_i\big((k+n-1)\cdot \Delta t\big) = \sum_{i}^{L} E_i \qquad (4)$$

where $M\cdot\Delta t \approx T_s$ is characteristic time (window) of the filter. After passing through the filter (averaging) the second term in (3) becomes negligibly small, because the signals of different paths are practically uncorrelated.

Thus, the signal energy at the output of analog part of the receiver increases with increasing number of paths. This leads to an increase of the ratio of energy-per-bit to the spectral density of noise $E_b/N_0$ ($N_0$ is determined by receiver parameters). Since in this scheme the receiver collects the energy of all incoming paths, it proves to be very effective.

We define ***multipath amplification of chaotic radio pulse*** as an increase of the pulse energy at the receiver input due to summation of the energy of pulses coming on a set of paths with respect to the energy of direct path in LOS (Line-Of-Sight) case, or with respect to the energy of the "strongest" path in NLOS (No Line-Of-Sight) case.

Whether this effect could be used to increase the receiver efficiency is determined by the type of the receiver. The receiver based on quadratic detector and LPF, which bandwidth matches the duration of chaotic radio pulse, can do this. It is an effective receiver of chaotic radio pulses in multipath channel.

In order to determine the effect of system parameters on multipath amplification, let us simulate DCC behavior in multipath environment using realistic UWB models of multipath channels.

## IV. CHANNEL MODELS FOR UWB SENSOR NETWORKS

In order to have a chance to compare different schemes for UWB communications, in the process of standardization of physical channel for UWB sensor networks (IEEE 802.15.4a) PHY Standard Committee developed a system of channel models for various environments. Nine models were developed, that describe signal propagation in 3.1–10.6 GHz range at distances 1–10 (30) m in the following media [19]: residential (CM-1/2), office (CM-3/4), outdoor (rural) (CM-5/6), industrial (CM-7/8) (all in LOS/NLOS versions), and for body area networks (range within 1 m). (В данной работе мы не будем рассматривать Body area networks.)

All models are statistical, i.e., each is based on a set of measurements made in a concrete environment. Besides the models, IEEE Committee defined unified rules of their use and also gave software routines for channel simulation that could be incorporated by researchers into their own simulation software. This gives designers possibility to estimate their system performance in equal conditions and simplifies comparison of simulation results.

This set of IEEE multipath channel models is based on two-scale Saleh-Valenzuela (S-V) model [20]. The signals at the receiver input are implied to come in relatively dense clusters characterized each by its own delay. Clusters contain paths with close delays. This two-level description allows us to account for propagation of electromagnetic waves in indoor environment where large uniform reflecting surfaces (walls, furniture, etc.) form groups of paths with similar characteristics. Paths' difference within cluster is due to peculiarities of concrete surface (nonuniform material, geometry, etc.).

UWB multipath channel is represented by a set of paths with random amplitude $\alpha_{k,l}$ and delay $\tau = T_l + \tau_{k,l}$ [19]. Channel response $h(t)$ is formed as a double sum of clusters and of paths within clusters:

$$h(t) = X \sum_{l=0}^{L}\sum_{k=0}^{K} \alpha_{k,l}\delta(t - T_l - \tau_{k,l}), \qquad (5)$$

where $\alpha_{k,l}$ is $k$th path gain within $l$th cluster; $\{T_l\}$ is arrival time of the first path of $l$th cluster; $\{\tau_{k,l}\}$ is delay of $k$th multipath component relative to $l$th cluster arrival time $T_l$; $X$ represents log-normal shadowing. Path statistics in $h(t)$ is described by Nakagami distribution with the following parameters: $\Lambda$ is cluster arrival rate; $\lambda$ is path arrival rate, i.e., the arrival rate of path within each cluster; $\Gamma$ is cluster energy decay rate; $\gamma$ is path energy decay rate within cluster.

Channel models CM-1 – CM-8 have different sets of parameters $\Gamma$, $\gamma$, $\Lambda$ and $\lambda$. In the next sections these models are used to estimate the effect of multipath amplification, and then to estimate a potential increase of communications range due to this effect.

## V. ESTIMATION OF MULTIPATH GAIN

As was defined above, multipath amplification is an increase of chaotic radio pulse energy at the receiver input due to multipath propagation. Numerically, multipath gain is the ratio of the energy of chaotic radio pulse after multipath channel to the energy of the same emitted pulse delivered by the "strongest" path.

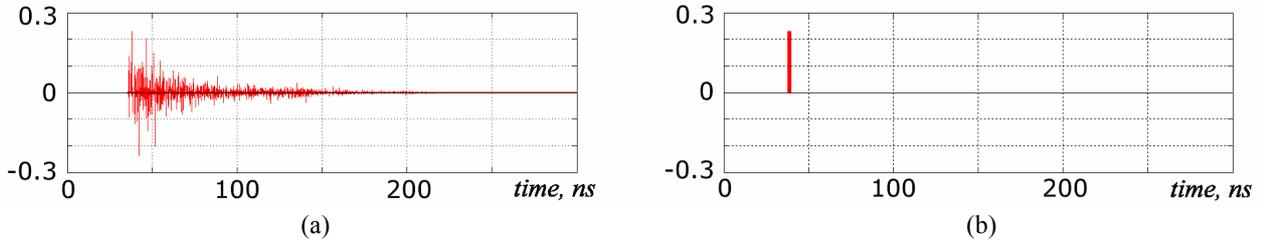

Fig. 6. (a) Multipath channel response; (b) single-path channel response

Let us estimate multipath gain in different channels. The procedure can be illustrated on the following example.

Let us take a NLOS channel for indoor residence (channel model CM-2 – Residential NLOS). A typical realization of channel response is shown in Fig. 6a. The strongest component of the response with amplitude $H_k \sim 0.23$ can be treated as the "main" path. If only this path is taken into account, the signal propagation is equivalent to passing through a single-path channel with response in Fig. 6b.

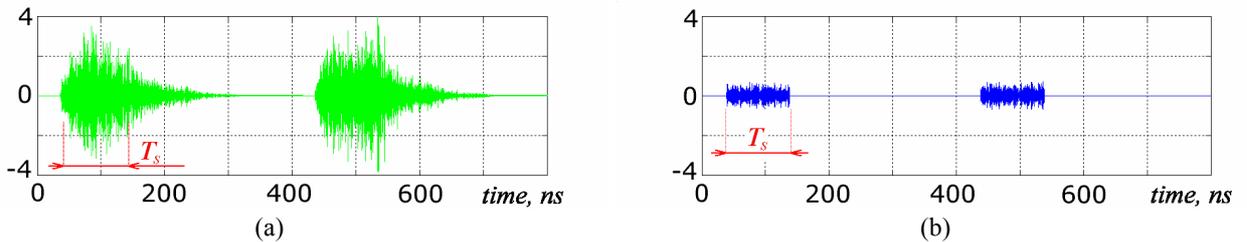

Fig. 7. Signal after (a) multipath and (b) single-path channels

To get the multipath gain estimate, we simulate transmission of a sequence of chaotic radio pulses of duration 100 ns (+ 300-ns guard interval) through multipath (Fig. 6a) and single-path (Fig. 6b) channels. This gives us signals in Figs. 7a and 7b, respectively. Then we integrate the energy of each pulse in the receiver on time interval $T_S$. In the case of multipath channel (Fig. 7a), some part of the energy is lost on guard interval. Multipath gain is calculated as the ratio of the energies of corresponding pulses in Figs. 7a and 7b. Then the values of multipath gain for this channel realization are averaged over many pulses.

In order to estimate multipath amplification effect in different conditions, we simulated signal propagation for all IEEE 802.15.4a standard channel models CM1–CM8. For each model 100 channel realizations were taken and 1000 chaotic radio pulses were put through each channel realization. Averaged results are given in Table 1.

As can be seen from Table 1, multipath gain is in the range 4 to 14 dB. In NLOS mode multipath gain is 0–10 dB higher than in LOS mode. This can be explained by the fact that in LOS mode most part of the energy is delivered by the main (first) path and contribution of other paths is relatively small. In NLOS mode, the path difference is smaller, so the multipath gain, i.e., contribution of the paths besides the "main" one, becomes more significant.

TABLE 1
MULTIPATH AMPLIFICATION GAIN IN DIFFERENT CHANNELS OF IEEE 802.15.4A STANDARD

| Environment | Multipath gain, dB | |
|---|---|---|
| | LOS | NLOS |
| Residence | 5 | 14 |
| Office | 4 | 12 |
| Open space | 5 | 5 |
| Industrial | 8 | 13 |

## VI. CONDITIONS NECESSARY FOR EFFECT OF MULTIPATH AMPLIFICATION ON COMMUNICATION SYSTEM PERFORMANCE

Thus, to observe multipath amplification effect, it is necessary that:
- radio pulse duration $T_S$ be larger than the time of autocorrelation $\tau$ of chaotic signal, and
- multipath channel delay $T_{delay}$ also be larger than autocorrelation time $\tau$.

If these conditions are satisfied, at least several paths on the time interval of channel delay $T_{delay}$ will be uncorrelated, thus, their energies will be summed.

However, these conditions are only necessary. From technical point of view, there are questions of communications efficiency, data rate, distance range and of the effect of multipath propagation on these parameters.

As was shown above, communication systems using chaotic signals are not subject to fading which is characteristic of narrowband carriers. However, to avoid inter-symbol interference, special measures must be taken, such as introducing guard interval.

An increase of average signal power at the receiver input due to multipath amplification leads, in the case of envelope detector, to an increase of pulse envelope at the output of analog part of the receiver, which gives better receiver performance. However, the pulse becomes longer due to delay spread and a part of it lays on guard interval. If the guard interval is short, the pulse can overlap with the next symbol position.

So, in order to exclude inter-symbol interference the duration of guard interval $T_g$ must exceed the multipath channel delay $T_{delay}$ or, at least, its part that carries the most (90–95%) portion of energy. In this case, the energy emitted during symbol transmission will not arrive at the position of next symbol (or only a negligibly small part of it will come, that will not significantly change error rate).

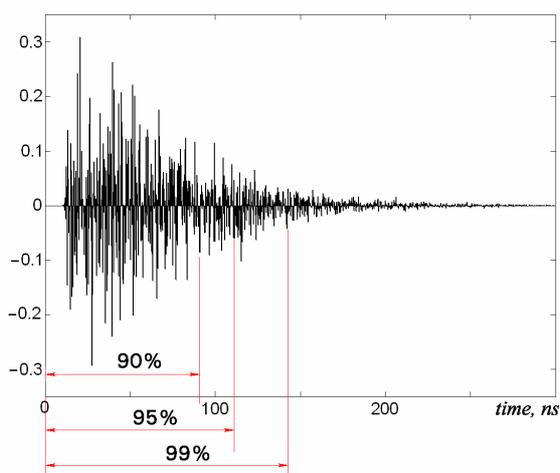

Fig. 8. Typical multipath channel response (CM4a-2, Residence NLOS).

When setting durations of chaotic radio pulses and of guard intervals, one must take into account autocorrelation time and channel delay as well as necessary information rate. Consider the following example.

Let the duration of multipath channel response $H_n$ be of the order of 150 ns and the radio pulse duration be 10 ns. Then, after the channel the pulse duration is extended to ~160 ns. To avoid the effect of inter-symbol interference, guard interval length must be set at no less than 100 ns (the main part of multipath response energy is located, as a rule, at its first third) (Fig. 8). From one hand, this imposes restrictions on the information rate. Here ($T_S$ = 10 ns, $T_g$ = 100 ns), it is $R$ = 1/110 ns ≈ 9 Mbps. From the other hand, if the receiver opens only on time interval of 10 ns equal to the original pulse duration, then the most part of pulse energy will be lost on guard interval (Fig. 5).

However, if the emitted radio pulse duration is 100 ns with the same 100-ns guard interval, then most part of the pulse energy will come at the pulse position and receiver efficiency increases. Yet the rate here is approximately twice lower: $R$ = 5 Mbps.

Consider the results of calculations of multipath gain *MG* as function of pulse duration (Fig. 9). The results for channel models CM-1 (Residence LOS) and CM-4 (Office NLOS) are presented. For each model several (4-6) typical channel realizations were taken. *MG* values were averaged over 2000 pulses.

As can be seen in Fig. 9, as the pulse duration decreases from $T_S$ = 100 ns, multipath gain decreases, as is expected, because the energy brought from delayed paths falls beyond information bit interval $T_S$, on which the energy is collected in the receiver. The rate of *MG* decrease is different not only in different channels, but also in different channel realizations within one channel model (i.e. Residence or Office). This can be explained by peculiar details of «energy profile» of corresponding multipath channel response. For instance, channel model CM-1 (Residence LOS) has shorter (more rapidly decaying) transient responses than model CM-4 (Office NLOS), so *MG* in CM-1 channel decreases approx. twice slower than in channel CM-4. When pulse duration $T_S$ is decreased from 100 to 40 ns, *MG* value in CM-4 decreases by 1–4 dB, whereas in CM-1 model only by 0.5–2 dB.

However, when chaotic pulse duration $T_S$ becomes essentially less than the duration of that part of the channel response that is responsible for transfer of the most part of energy, *MG* value drops rapidly (in Fig. 9 this corresponds to $T_S$ = 20–30 ns).

As follows from this analysis, multipath amplification is observed with sufficiently long pulses. For short pulses there is no multipath amplification.

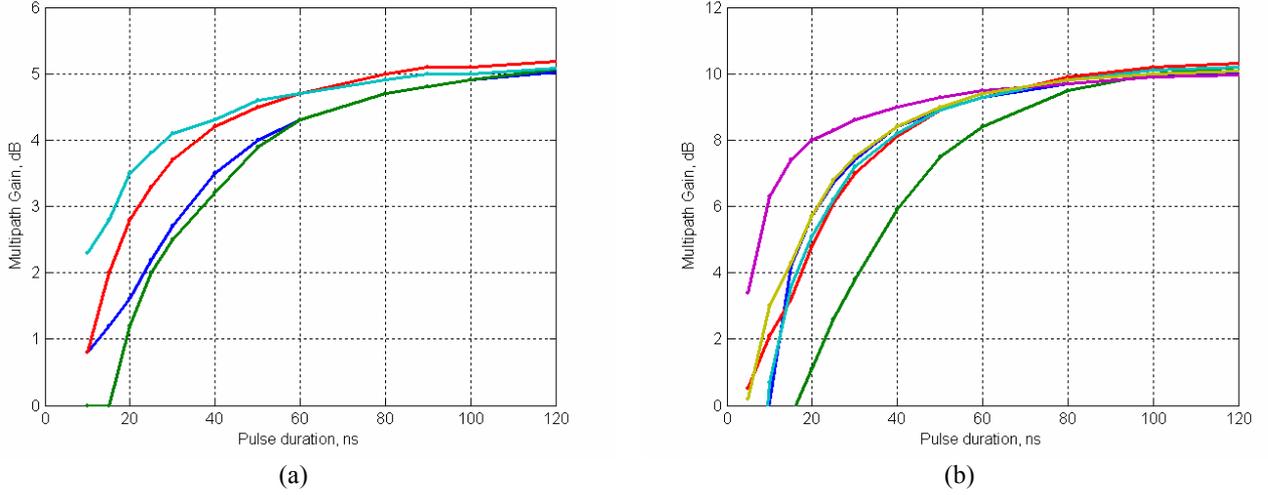

Fig. 9. Multipath gain as function of duration of chaotic radio pulse: (a) channel model CM-1 (Residence LOS), (b) CM-4 (Office NLOS)

With conditions on existence of multipath amplification effect satisfied and no inter-symbol interference, communication system efficiency increases (in respect to $E_b/N_0$ at the receiver input) with increasing chaotic pulse duration. However, this is achieved on account of lower transmission rate.

## VII. EFFECT OF MULTIPATH AMPLIFICATION ON PERFORMANCE OF WIRELESS UWB SENSOR NETWORK

Evidently, multipath amplification effect can increase operation range of communications system. The effect is more profound in the case of longer pulses, so let us consider a relatively low-rate communication system of abovementioned IEEE 802.15.4a standard intended for UWB wireless sensor networks. According to the standard, such networks can include many thousand sensors that from time to time transmit small amount of information. Nodes of such networks must be supplied with UWB (bandwidth 0.5 to 7.5 GHz) transceivers of frequency range 3.1–10.6 GHz with rates 1–1000 kbps. Special requirements to the transceivers are: very low emission level (maximum average spectral density –41.3 dBm/MHz), low power consumption (batteries must work for at least 2 years) and low cost (less than $1).

Consider a direct chaotic communication system complying the requirements of IEEE 802.15.4a standard. The parameters of DCC system are as follows:
- frequency range $f$ = 3.1–5.1 GHz;
- signal bandwidth $\Delta f$ = 2 GHz;
- average transmitter power (emitted) $P_{tx} = S\Delta f$ = –41.3 dBm/MHz × 2000 MHz = –8.3 dBm = 0.15 mW;
- transmission rate $R$ = 1000 kbps;
- transmitter antenna gain $G_{tx}$ = 0 dB;
- receiver antenna gain $G_{rx}$ = –3 dB;
- noise factor $NF$ = 7 dB;
- implementation loss $I$ = 3 dB;
- BER = $4 \cdot 10^{-5}$ (PER = $10^{-2}$);
- minimum value of $E_b/N_0$ = 18.5 dB in free space, 19.5–21 dB in multipath channels.

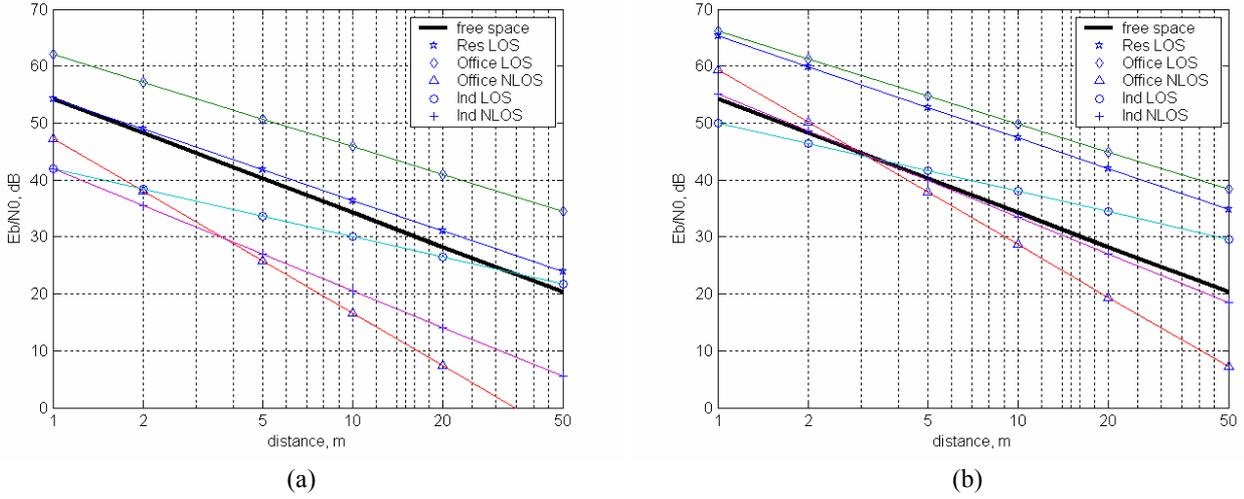

Fig. 10. $E_b/N_0$ at the receiver input as a function of distance in (a) the absence of multipath amplification and (b) with multipath amplification

The standard requires operation distance no less than 30 m (100 m optional).

Operation range of DCC system is the distance at which the energy-per-bit to spectral-density-of-noise ratio $E_b/N_0$ decreases down to the minimum. The value of this minimum depends on the receiver structure, on modulation method, etc. As is stated above, the prescribed value of BER = $4\cdot10^{-5}$ is achieved with DCC systems in free-space channel at $E_b/N_0$ = 18.5 dB, whereas in different multipath channels the same BER value is obtained at $E_b/N_0$ in the range 19.5–21 dB.

Thus, in order to determine the system range, for each channel we must plot $E_b/N_0$ as function of distance $d$ and determine the distance at which the value of $E_b/N_0$ reaches corresponding limit value.

Let us calculate $E_b/N_0$ from the signal power at receiver input.

$$E_b/N_0 = (P_{rx} \cdot T_b) / kT = P_{rx} / (kT \cdot R), \qquad (6)$$

where $P_{rx}$ is the power of the receiver input signal, $T_b$ is information bit duration, $T_b = 1/R$ ($R$ is transmission rate), $N_0 = kT$ is the spectral density of noise.

To calculate $P_{rx}$, we use ordinary Link Budget calculus:

$$P_{rx} = P_{tx} G_{tx} G_{rx} / PL(d)\ I\ NF, \qquad (7)$$

or in notation, more usual in radio engineering:

$$P_{rx}\ [dBm] = P_{tx}\ [dBm] + G_{tx}\ [dB] + G_{rx}\ [dB] - PL(d)\ [dB] - I\ [dB] - NF\ [dB], \qquad (8)$$

where $P_{tx}$ is transmitter power, $G_{tx}$, $G_{tx}$ is antenna gain of transmitter and receiver respectively, $PL(d)$ is path loss, i.e. signal attenuation at distance $d$ from transmitter, $I$ is implementation loss, and $NF$ is noise factor (noise added by receiver).

When considering transmitter power $P_{tx}$, one must take into account the signal structure. Note that the above power $P_{tx}$ = 0.15 mW (–8.3 dBm) is average, so the real power is twice as high (plus 3 dB), because when symbol "0" is transmitted, no power is emitted, and symbols "0" and "1" are equally probable, i.e. $P_{tx}$ = 2<$P_{tx}$>, where <> means averaging. Besides, guard intervals (duty cycle) must be also accounted for, so $P_{tx}$ = 2<$P_{tx}$>$T_b/T_s$, where $T_b$ is the whole bit duration and $T_s$ is chaotic radio pulse duration.

In the discussed channel models, signal attenuation on the path from transmitter to receiver (Path Loss) $PL(d)$ is described by power function of distance $d$ [19]

$$PL(d) = PL(d_0) \cdot 1/(d/d_0)^n, \qquad (9)$$

where $PL(d_0)$ is path loss at reference distance $d_0 = 1$ m and power $n$ determines the rate of signal attenuation (in free space $n = 2$). Parameter $PL_0$ is also defined by channel model and it depends on the environment (see Table 2).

TABLE 2
MAXIMUM RANGE OF DCC SYSTEM IN MULTIPATH ENVIRONMENT

| Channel model | $N$ | Attenuation $PL_0$ at distance 1 m, dB | Range without multipath gain, m | Range with multipath gain, m |
| --- | --- | --- | --- | --- |
| Free space | 2 | 44.4 | 33 | 33 |
| Residence, LOS | 1.79 | 44.4 | 49 | 203 |
| Residence, NLOS | 4.58 | 44.4 | 5 | 9 |
| Office, LOS | 1.63 | 36.6 | 218 | 384 |
| Office, NLOS | 3.07 | 51.4 | 6 | 14 |
| Open space, LOS | 1.76 | 43.3 | 61 | 117 |
| Open space, NLOS | 2.5 | 43.3 | 18 | 38 |
| Industrial, LOS | 1.2 | 56.7 | 32 | 147 |
| Industrial, NLOS | 2.15 | 56.7 | 7 | 28 |

Thus, the equation for $E_b/N_0$ is as follows

$$E_b/N_0 = P_{tx}\ G_{tx}\ G_{rx} / PL_0\ d^n\ I\ NF\ kT\ R. \qquad (10)$$

In Fig. 10a the value of $E_b/N_0$ is shown as a function of the distance between transmitter and receiver. For comparison, the case of free space is depicted with solid line. Note that NLOS channels are characterized by much higher attenuation exponent $n > 3$ than LOS channels ($n < 2$) (see Table 2). Note also that reference attenuation $PL_0$ at a distance of 1 m in NLOS channels is by 0–15 dB larger than that for corresponding LOS channels. Consequently, operation range in NLOS channels, as a rule, is significantly shorter with respect to free space.

To determine operation range of DCC system using this plot, one must find the distance at which $E_b/N_0$ decreases down to minimum admissible value $(E_b/N_0)_{min}$, at which BER (bit-error ratio) decreases to a predetermined value, say, $P = 4 \cdot 10^{-5}$ (which corresponds to packet-error ratio PER = $10^{-2}$). For DCC system in free space $(E_b/N_0)_{min} = 18.5$ dB, whereas in multipath environments $(E_b/N_0)_{min} = 19.5–21$ dB depending on the environment (residence, office, etc.). Operation range estimates are given in Table 2.

The effect of multipath amplification improves the situation drastically. In order to take it into account, Eq. (8) must be rewritten as

$$E_b/N_0 = P_{tx}\ G_{tx}\ G_{rx} / PL_0\ d^n\ I\ NF\ kT\ R\ MG. \qquad (11)$$

In Fig. 10b the dependence of $E_b/N_0$ on distance $d$ is shown with multipath amplification taken into account. Distance estimates are also stored in Table 2. As can be seen from this table, the effect of multipath gain gives much larger distance range.

With two exceptions (Residence and Office NLOS), the system range complies requirements of IEEE 802.15.4a. As follows from Eq. (11), 30-m range can be achieved in these channels by means of decreasing transmission rate down to 4 and 100 Kbps, respectively.

Thus, use of UWB chaotic pulses as information carrier and envelope detector as receiver allows us to solve certain problems of multipath propagation (fading, inter-symbol interference), and to increase the system range due to effect of multipath amplification.

## VIII. CONCLUSIONS

Analysis of performance of direct chaotic communication system in multipath channel showed that naturally wideband character of chaotic oscillations eliminates the problem of signal fading.

Multipath amplification effect is found, which, for a proper relation between channel delay, durations of chaotic radio pulses and guard intervals, can essentially increase effective useful power at the receiver input.

Envelope detector matched by structure of information signal is shown to be an effective receiver of communication system with chaotic radio pulses in multipath channels.

Multipath gain was estimated for realistic UWB models of multipath channels derived by IEEE Standardization Committee for IEEE 802.15.4a standard. Multipath amplification gives 5–15 dB gain in energy efficiency of the system depending of the type of multipath environment, which allows, e.g., to increase operation range by a factor of 2–5 at constant average transmitted power.

**Yuri V. Andreyev** (b. 1960) graduated from the Moscow Institute of Physics and Technology in 1983. Received Ph.D. degree from the Institute of Radio Engineering and Electronics (IRE) of Russian Academy of Sciences in 1993.

Since 1983 he is with the IRE, now working in the field of information and communication technologies based on chaotic dynamics.

**Alexander S. Dmitriev** (M'97) (b. 1948) graduated from the Moscow Institute of Physics and Technology in 1971. Received Ph.D. degree in 1974 from the same Institute. In 1988 he received Dr.Sci. degree from the Institute of Radio Engineering and Electronics (IRE) of Russian Academy of Sciences, Moscow.

At present, he is the chief of InformChaos Lab. in IRE. His research interests include nonlinear dynamics, communication and information technologies, and chaos.

**Andrey V. Kletsov** received the M.Sc. degree in applied physics and mathematics from the Moscow Institute of Physics and Technology in 2005. Now he is a Ph.D. student in the Institute of Radio Engineering and Electronics (IRE), Moscow, and is working on his thesis in the InformChaos Lab.